\documentclass{aa}
\usepackage{psfig} 

\def\ros{{\sl ROSAT }}
\def\asca{{\sl ASCA }}

\def\ein{{\sl Einstein }}

\def\G{$\Gamma_{\rm x}$ }
\def\NH{$N_{\rm H}$ }

\def\approxlt{\mathrel{\hbox{\rlap{\lower.55ex \hbox {$\sim$}}
        \kern-.3em \raise.4ex \hbox{$<$}}}}
\def\approxgt{\mathrel{\hbox{\rlap{\lower.55ex \hbox {$\sim$}}
        \kern-.3em \raise.4ex \hbox{$>$}}}}

\begin{document}

  \thesaurus{03         
             (11.01.2;  
               11.06.2;  
               11.09.4;  
               11.14.1;  
               13.25.2)  
}

   \title{X-ray properties of LINERs}  
   \author{ Stefanie Komossa\inst{1}, Hans B{\"o}hringer\inst{1}, John P. Huchra\inst{2}}
\offprints{St. Komossa,\\
 skomossa@xray.mpe.mpg.de}
\institute{
Max-Planck-Institut f\"ur extraterrestrische Physik,
         Postfach 1603, D-85740 Garching, Germany
\and
Harvard-Smithsonian Center for Astrophysics, 60 Garden St., MS-20, Cambridge, MA02138, USA}
\date{Received: 6 April 1999; accepted: 17 June 1999 }
   \maketitle
\markboth{St.~Komossa et al.: X-ray properties of LINERs}
{St.~Komossa et al.: X-ray properties of LINERs}

   \begin{abstract}
We present an investigation of the X-ray properties of 13 LINERs
based on \ros all-sky survey and pointed PSPC and HRI observations. 
Several sources are studied for the first time in X-rays.   
The X-ray spectra are best described by a powerlaw with photon index
\G $\approx$ --2.0 or thermal emission from gas with very subsolar 
abundances.  
The luminosities range between $\log L_{\rm x} = 37.7$ (NGC\,404)
and 40.8 (NGC\,4450).        
No X-ray variability on the timescale of hours/days is detected.
This is in line with the suggestion that LINERs may accrete in 
the advection-dominated mode.
On a longer term, one of the objects, NGC\,2768, turns out to
be slightly variable. 
Some sources appear to be extended at weak emission levels whereas the bulk of the X-ray
emission is consistent with arising from a point source within the PSPC instrumental resolution.  
$L_{\rm x}$/$L_{\rm B}$ ratios are derived and emission mechanisms 
that potentially contribute to the observed X-ray luminosities are discussed. 

We also examine the presence of second X-ray sources near the target
sources both in terms of instrumental effects and in terms of `real' sources 
within the LINER galaxies.

   \end{abstract}
%
\section{Introduction}

Low-ionization nuclear emission line regions, LINERs,
are characterized by their optical emission line spectrum
which shows a lower degree of ionization
than Seyfert galaxies (e.g., Heckman et al. 1980). 
Their major power source and line excitation mechanism 
has been a subject of lively debate ever since their discovery
(for reviews see, e.g., 
Filippenko 1989, 1993, Ho 1998). 
LINERs manifest the most common type of activity
in the local universe.
If powered by accretion, they probably represent the low-luminosity end
of the quasar phenomenon 
and their presence has relevance to, e.g., the evolution of quasars, 
the faint end of the AGN luminosity function, 
and the presence of supermassive black holes (SMBHs) in nearby galaxies. 
A detailed study
of the LINER phenomenon is thus very important. 

Many different mechanisms that might account for their optical emission line
spectra have been examined, including 
collisional ionization and excitation (Burbidge \& Burbidge 1962), 
shock heating (e.g., Fosbury
et al. 1978, Heckman 1980, Dopita et al. 1996, Contini 1997),  
photoionization by hot
stars (e.g., Shields 1992, Ho et al. 1993), photoionization
by a non-stellar continuum source (e.g., Ferland \& Netzer 1983,
Halpern \& Steiner 1983, Binette 1984,1985,1986, Ho et al. 1993),  
and photoionization by an 
absorption-diluted AGN continuum (Halpern \& Steiner 1983,
Schulz \& Fritsch 1994).
Despite this detailed shock and photoionization modelling 
the nature of the main ionizing source of LINERs
remained elusive, although there is now growing evidence that 
they are accretion powered (e.g., Falcke et al. 1997, Falcke 1998, Ho 1998). 
Eracleous et al. (1995), in an effort to explain the UV bright 
centers detected in some but not all LINERs, suggested
a duty cycle model where central activity in LINERs is governed by
occasional tidal disruptions of stars by central black holes.
In an alternative approach, Barth et al. (1998) suggested that
dust extinction could cause the UV darkness of some LINERs.  

X-rays are a powerful tool to investigate the presence of an AGN 
via X-ray variability, luminosity, and extent, and to explore the physical properties
of LINERs in general. 
Nevertheless, not many LINERs have been examined in X-rays, particularly not
larger samples in an homogeneous way. The largest previous one we are
aware of was presented by Ptak et al. (1999, see also Serlemitsos et al. 1997)
and consisted of several low-luminosity AGN (LLAGN) including 5 LINERs
observed with {\sl ASCA}. 
They find that the \asca spectra are best described by 
a two-component model consisting of soft thermal emission  
and a powerlaw with photon index \G $\approx -1.7$, with varying relative
contributions of the two spectral components from object to object. 
Several studies of individual objects 
(e.g., Mushotzky 1982, Koratkar et al. 1995, Ehle et al. 1995,
Cui et al. 1997, Terashima et al. 1998, Pietsch et al. 1998, Roberts et al. 1999)
revealed
results consistent with the above spectral results. 
A fairly complex X-ray spectrum was recently reported for the LINER 
NGC\,1052 (Weaver et al. 1999).
Concerning the spatial extent of the X-ray emission,
Koratkar et al. (1995; K95 hereafter) found their two LINERs to
be consistent with a point source within the limits of the \ros HRI
resolution.
X-ray luminosities range up to $\sim$ 10$^{41}$ erg s$^{-1}$ (e.g., 
Ptak et al. 1999, Stockdale et al. 1998, K95; see Halpern \& Steiner 1983 for
a collection of \ein luminosities) and are presently biased towards the X-ray brightest
objects. 

Given the importance to better understand the LINER phenomenon
and activity in nearby galaxies in general,
with its potential bearing on the evolution of SMBHs in galaxies,
the contribution to the faint end of the AGN luminosity function, and the
soft X-ray background;
and given the still limited number
of objects previously studied in the X-ray spectral region, 
we examined a sample of 13 LINERs with the \ros (Tr\"umper 1983) instruments
(Pfeffermann et al. 1987). 
We report here the 
results of an investigation of the spectral, spatial, and temporal
X-ray properties of these galaxies.    

Our sample consists of spiral galaxies and lenticulars.
The primary selection was according to the list of 
Huchra \& Burg (1992, with most LINERs identified in Heckman 1980,
Staufer 1982, and Keel 1983) and Huchra (1998, priv. com.). We then excluded
LINERs that were, on the basis of emission lines, re-classified
as Seyferts or listed
to contain a Seyfert-component as well according to the NED database. 
This resulted in 13 remaining LINERs.
Results for a larger sample of LLAGN, including the objects with composite
spectra and Seyfert\,2 galaxies will be reported elsewhere. 
For the selected sources we analyzed
\ros all-sky survey data. In addition, 8 of the galaxies were targets of, or serendipituously
located in the field of view of, PSPC and/or HRI observations. 
All except one are detected,
and for 5 of them a PSPC spectral analysis was possible. 
The brightest source turned out to be NGC\,4450 which is studied in most detail below.   

The paper is organized as follows: 
The data reduction is described in Sect. 2. In the next two
Sections we present the general assumptions on which the data analysis
is based (Sect. 3) and results for the individual objects (Sect. 4),
including a discussion of (the reality of) further X-ray sources
close to the target sources.  The discussion (Sect. 5) is followed
by the concluding summary in Sect. 6.

\section{Data reduction} 

We used \ros all-sky survey (RASS) data as well as
archival and serendipituous pointed observations of the galaxies. 
The observations are summarized in Table 1.

For further analysis of the pointed PSPC and HRI data
the source photons were extracted
within a circular cell centered on the target source. 
The background was determined in a source-free region around 
the target source and subtracted. 
The data were corrected for vignetting 
and dead-time, using the EXSAS software package (Zimmermann et al. 1994).

To carry out the spectral analysis source photons in
the amplitude channels 11-240 were binned
according to a constant signal/noise ratio of $>$ 4$\sigma$.  

Lightcurves were created with a binning of 800 sec to account for
the wobble mode of the satellite. 

   \begin{table*}             
     \caption{Log of observations. $t_{\rm exp}$ gives the exposure time. 
       If `surv' is listed in the column `instrument' this refers to RASS data
       taken with the PSPC. RASS observations were performed during Aug. 1990 -- Jan. 1991.   
       }
     \label{obslog}
      \begin{tabular}{llcrc}
      \hline
      \noalign{\smallskip}
        galaxy & date &  instrument & $t_{\rm exp}$ & $CR$$^{(1)}$ \\  
       \noalign{\smallskip}
             &        &             &  sec          & cts/s     \\ 
      \noalign{\smallskip}
      \hline
      \hline
      \noalign{\smallskip}
  NGC\,404  &           & surv & 473 & $<$0.02  \\ 
            & 04/01/97 & HRI     & 23874 & 0.0011 $\pm$0.0003 \\ 
\noalign{\smallskip}
 NGC\,1167  &           & surv & 477 & $<$0.02 \\ 
            & 27-28/1/92 & PSPC & 7315 & $<$0.005 \\ 
\noalign{\smallskip}
 NGC\,2768  &           & surv & 341 &  0.038 $\pm$0.014 \\ 
            & 11-13/10/93 & PSPC & 4766 & 0.021 $\pm$0.002 \\ 
            & 01/05/94    & PSPC & 2898 & 0.013 $\pm$0.003 \\ 
\noalign{\smallskip}
 NGC\,3642  &           & surv & 540 &  0.023 $\pm$0.011 \\ 
            & 10-11/5/93 & PSPC & 6178 & 0.018 $\pm$0.003 \\ 
            & 21/10/93   & PSPC & 7804 & 0.016 $\pm$0.002 \\  
\noalign{\smallskip}
 NGC\,3898  &           & surv & 456 & 0.016 $\pm$0.011 \\ 
            & 19/11/91  & PSPC & 4916 & 0.012 $\pm$0.003 \\ 
\noalign{\smallskip}
 NGC\,4036  &           & surv & 415 & 0.013 $\pm$0.010 \\ 
\noalign{\smallskip}
 NGC\,4419  &           & surv & 73 & $<$ 0.14 \\ 
\noalign{\smallskip}
 NGC\,4450  &           & surv & 266 & 0.039 $\pm$0.018 \\ 
            & 04-12/6/92 & PSPC & 6170 & 0.101 $\pm$0.004 \\ 
            & 17/12/92   & PSPC & 5145 & 0.091 $\pm$0.004 \\ 
\noalign{\smallskip}
 NGC\,5371  &           & surv & 682 &    0.026 $\pm$0.010 \\ 
            & 23/06/ -01/07/93 & PSPC & 14960 & 0.038 $\pm$0.002 \\    
\noalign{\smallskip}
 NGC\,5675  &           & surv & 623 & $<$0.016 \\ 
\noalign{\smallskip}
 NGC\,5851  &           & surv & 123 & $<$0.08 \\ 
\noalign{\smallskip}
 NGC\,6500  &           & surv & 262 & $<$0.04 \\ 
            & 17-20/3/93 & HRI   & 5953 & 0.0025 $\pm$0.0008 \\ 
\noalign{\smallskip}
 IC\,1481  &            & surv & 410 & $<$0.02 \\ 
\noalign{\smallskip}
      \noalign{\smallskip}
      \hline
      \noalign{\smallskip}
  \end{tabular}

\noindent{$^{(1)}$countrate in the observed (0.1-2.4) keV band (note the different
sensitivity of the PSPC and HRI detector, which results in a countrate conversion 
of a factor of several from one instrument to the other, depending
on the spectral shape). 
}
   \end{table*}

\section {Data analysis: General considerations and assumptions}
The following models were fit to the X-ray spectra of all objects: (i) a powerlaw 
of the form $\Phi \propto E^{\Gamma_{\rm x}}$, and (ii) emission from a Raymond-Smith
(1977; RS hereafter) plasma, abundances fixed to either the solar value (Anders \& Grevesse 1989)
or below, up to 1/100 $\times$ solar. The amount of cold absorption was constrained not
to underpredict the Galactic value (Dickey \& Lockman 1990) along the line-of-sight. 

If several PSPC observations were available, we used the one with the deepest exposure
time (for the present sample, this also always happened to correspond 
to the pointing were the source was on-axis, if an on-axis pointing existed at all.) 

To calculate X-ray fluxes and luminosities, we proceeded as follows:  
For sources bright enough to allow spectral fits
we integrated over the SED in the (0.1-2.4 keV) band after correcting
for cold absorption. For RASS sources too weak to perform
spectral fits, and for HRI sources we assumed a powerlaw spectrum  
with \G=--1.9 and absorption of the Galactic value. 
Distances were calculated using a
Hubble constant of $H_0$=75 km/s/Mpc for the distant ($v > 3000$ km/s) objects.
For those nearby or even with blueshift we used Tully's (1988) catalog
of nearby galaxies (if not stated otherwise) which is based on the virgocentric model
of Tully \& Shaya (1984). 
To check for the influence of the assumed distances, which can be
important  given that all galaxies are very nearby, we re-calculated
all luminosities based on distances obtained with the flow field model
of Mould et al. (2000; ApJ, in prep.). We find that our conclusions are unaltered and that
luminosities of individual objects are changed by a factor $\approxlt$ 2. 
If not stated otherwise, X-ray luminosities given below
refer to the energy interval (0.1--2.4) keV.  

Some of the RASS sources were not significantly detected. Namely,
formally only 1, 3, 0, 5, 1, 2  source photons above the background were 
registered in the (0.5--2) keV band  
for NGC\,404, NGC\,1167, NGC\,4419, NGC\,5675, NGC\,5851, and IC\,1481, respectively. 
In this case we conservatively assumed
that $<$10 source photons would have escaped detection 
and the upper limits for countrates and luminosities
listed in Tables 1, 2 were calculated correspondingly.   

To derive blue luminosities, we used the observed blue magnitudes of 
de Vaucouleur et al. (1991; see also Huchra \& Burg 1992). 
To carry out the extinction correction we adopted the same amount of
Galactic absorption (Dickey \& Lockman 1990) as we did in the X-ray analysis,
assumed a standard gas/dust ratio, and utilized the relation of 
Bohlin et al. (1978; see also Predehl \& Schmitt 1995).

  \begin{figure*}[t]
      \vbox{\psfig{figure=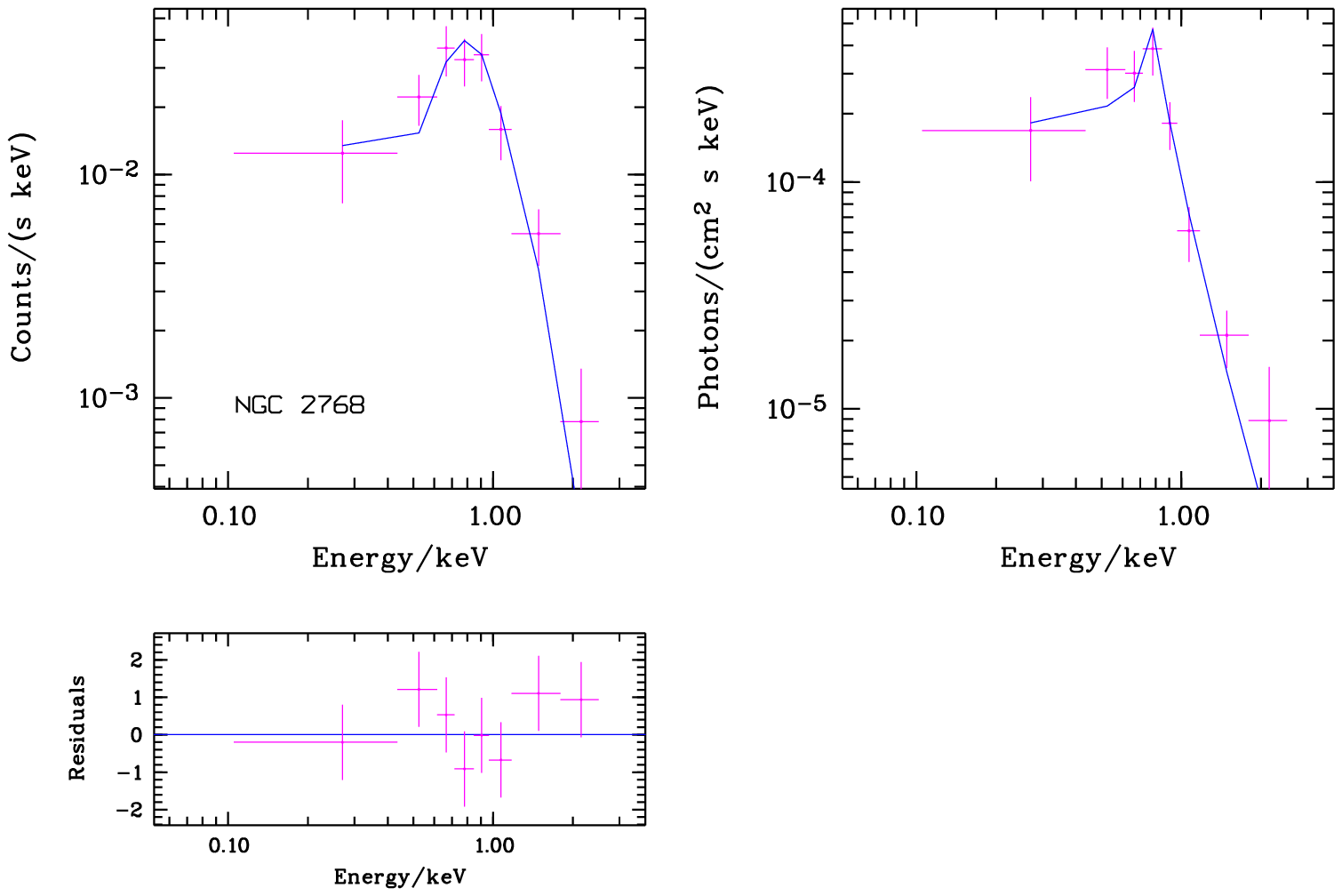,width=4.9cm,height=7.0cm,%
          bbllx=2.1cm,bblly=1.1cm,bburx=10.1cm,bbury=11.7cm,clip=}}\par
       \vspace*{-7.0cm}\hspace*{4.7cm}
      \vbox{\psfig{figure=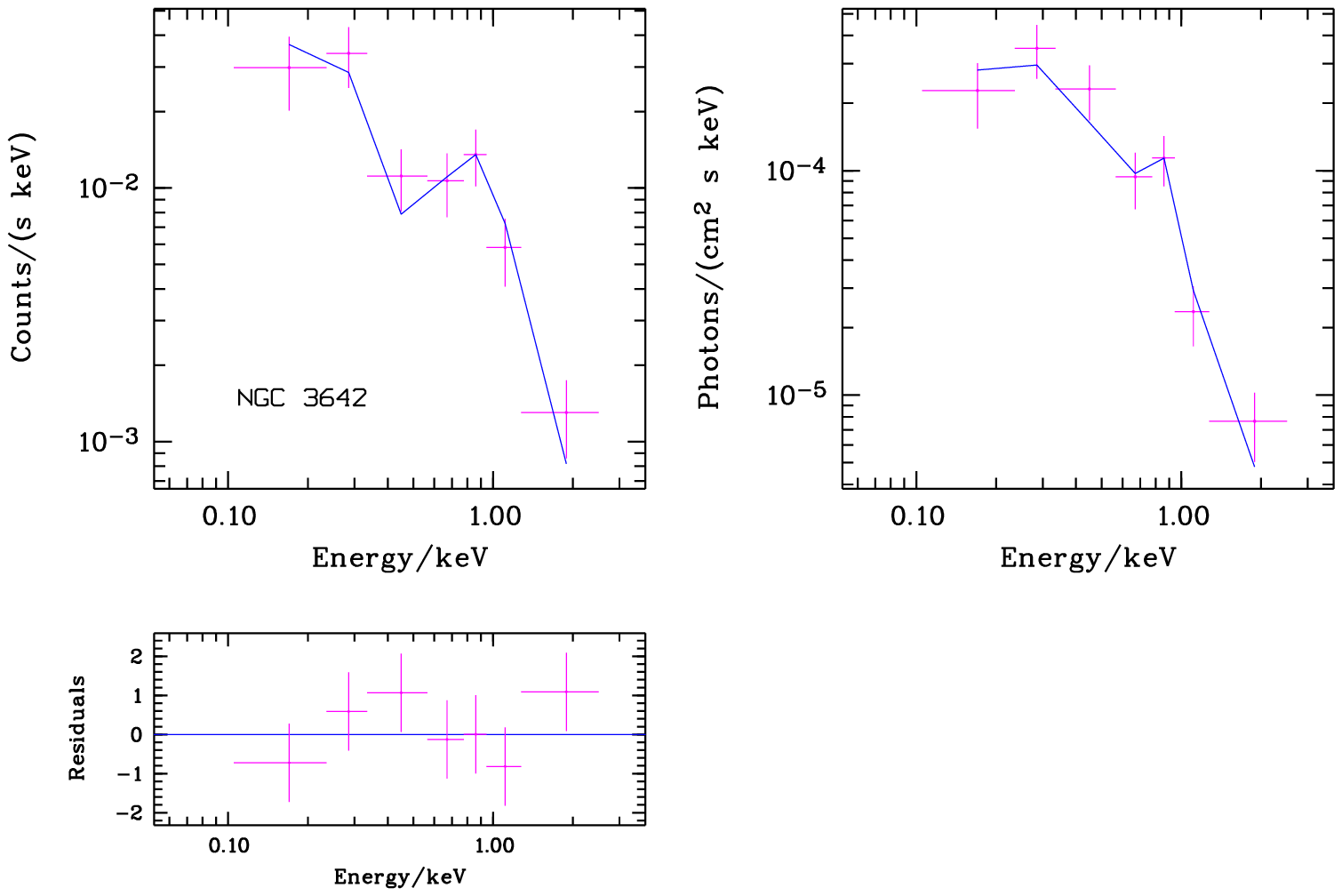,width=4.4cm,height=7.0cm,%
          bbllx=3.0cm,bblly=1.1cm,bburx=10.1cm,bbury=11.7cm,clip=}}\par
       \vspace*{-7.0cm}\hspace*{9.0cm}
      \vbox{\psfig{figure=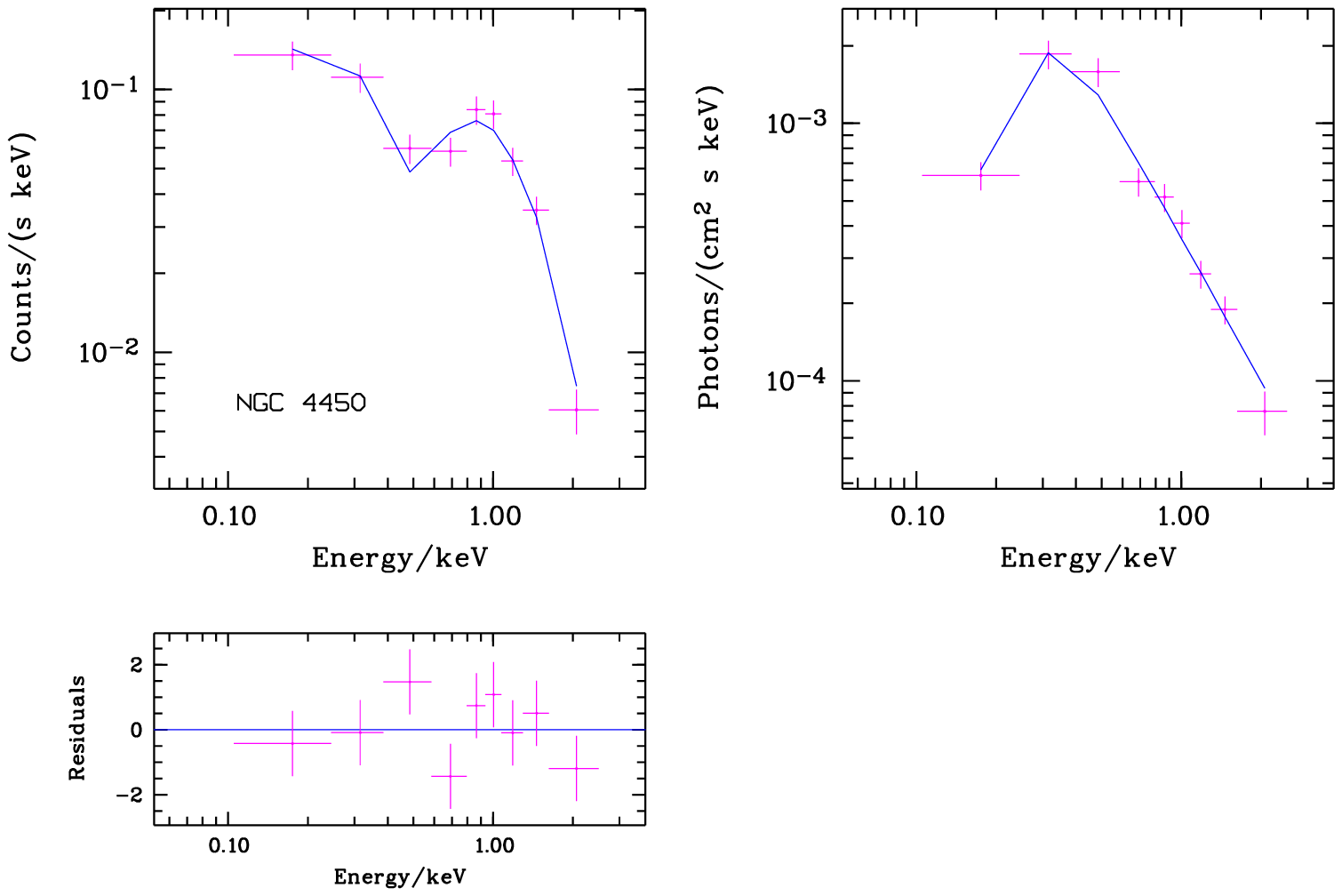,width=4.4cm,height=7.0cm,%
          bbllx=3.0cm,bblly=1.1cm,bburx=10.1cm,bbury=11.7cm,clip=}}\par
       \vspace*{-7.0cm}\hspace*{13.5cm}
      \vbox{\psfig{figure=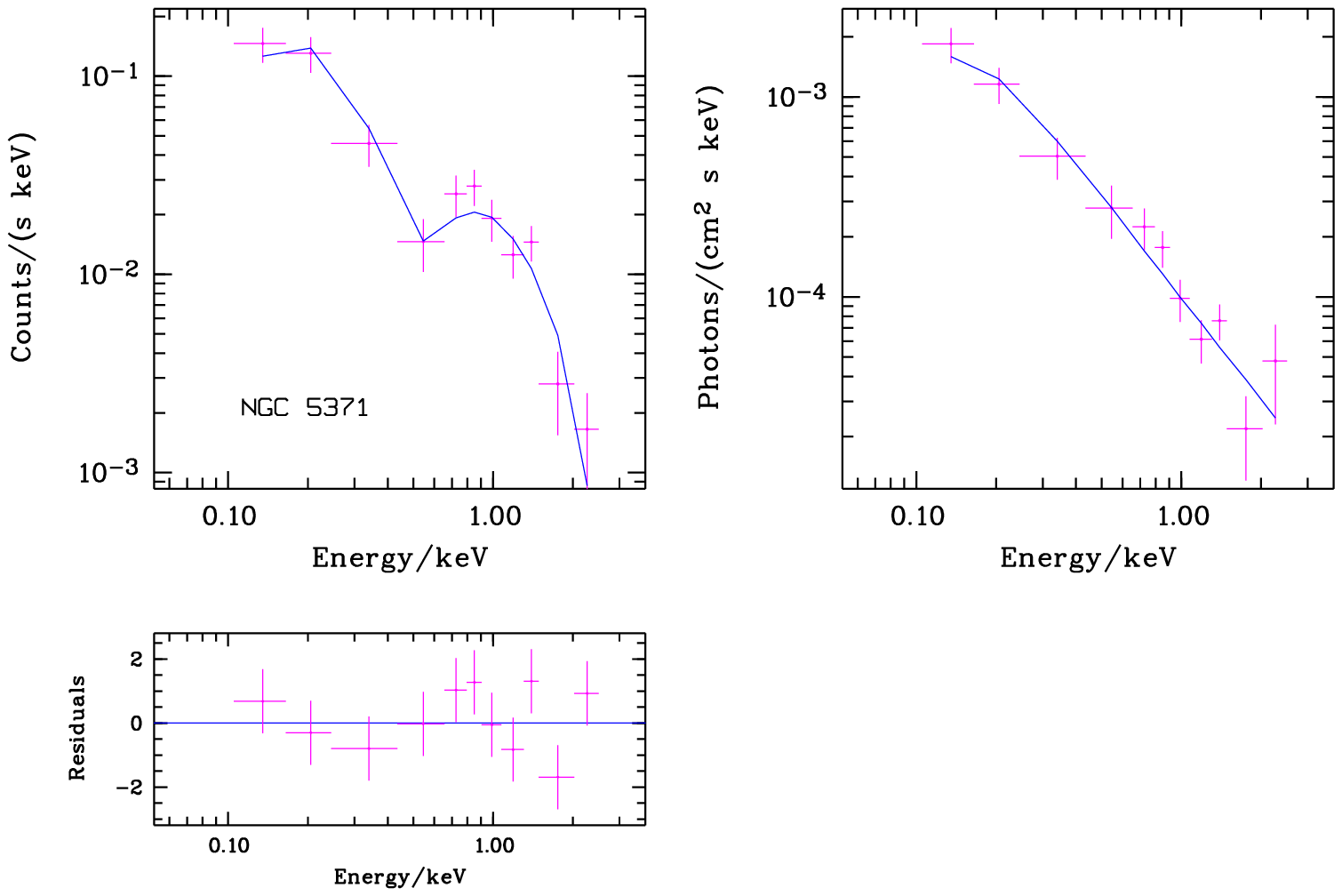,width=4.4cm,height=7.0cm,%
          bbllx=3.0cm,bblly=1.1cm,bburx=10.1cm,bbury=11.7cm,clip=}}\par
 \caption[spec]{ X-ray spectra and fit residuals for the four brightest PSPC-observed
sources. The upper panel gives the observed X-ray spectrum of each galaxy 
(crosses) and the model fit (solid line). The lower panel shows the fit residuals. 
NGC\,2768: RS fit with abundances of 0.1\,solar; 
NGC\,3642: RS fit with abundances of 0.05\,solar;
NGC\,4450: pl fit; NGC\,5371: pl fit (see Table 2 for details).  
}
\label{spec}
\end{figure*}

\section {Notes on individual objects} 
Below, we first give a brief summary of what is known for the individual
galaxies (only the detected ones plus NGC\,1167)
from the literature 
and then describe the results from our X-ray temporal, spectral and spatial
analysis of the individual objects.  

A detailed literature search revealed that some of the present
galaxies were already very briefly discussed 
in other/previous samples with different aims. 
Given the inhomogeneity of the assumptions made and models fit
(for details see below),
we extent here the spectral analysis of these objects and also perform a spatial
and temporal analysis. 

\subsection {NGC\,404}
NGC\,404 is blueshifted (Stromberg 1925). 
Baars \& Wendker (1976) noted its peculiar radio properties.  
Optical spectroscopy was performed by, e.g., Burbidge \& Burbidge (1965),
Keel (1983), and Filippenko \& Sargent (1985) and revealed very narrow emission
lines;
for an image see Sandage (1961). 
Larkin et al. (1998) obtained NIR spectra and reported the detection of strong
[FeII] emission in this and several further LINERs (but not in all
of their sample) and suggested X-ray heating to be at work. 
A molecular gas ring was observed by Wiklind \& Henkel (1990).
The detection of a UV core with HST was presented  
by Maoz et al. (1995). Based on the analysis of UV spectra,
Maoz et al. (1998) explained the data by the presence of a 
central star cluster.  

A deep HRI observation of NGC\,404 is available. The source is detected ($\sim$25 source
photons) but too weak to allow a more detailed temporal or spatial analysis.  
Assuming a powerlaw spectral shape as described above we derive a (0.1-2.4 keV)
luminosity of $L_{\rm x} = 10^{37.7}$ erg s$^{-1}$, the lowest $L_{\rm x}$ among the
present objects, and among the lowest so far detected for a LINER.
The X-ray emission of NGC\,404 is consistent with originating completely from
discrete stellar sources given the galaxy's blue luminosity. Using the relation
between $L_{\rm x}$ and $L_{\rm B}$ of Canizares et al. (1987), 
we predict $L_{\rm x}^{0.5-4.5 \rm keV} = 10^{38.25}$ erg s$^{-1}$ in the (0.5--4.5 keV) band
 which compares to the observed value of  
$10^{37.6}$ erg s$^{-1}$, which is below the expectation but consistent within
the scatter. (The intrinsic X-ray luminosity could be boosted
if there is some excess absorption or the spectral shape is different from the
assumed one.) It is interesting to note that Wiklind \& Henkel (1990) argue for a much 
larger distance of NGC\,404 than derived from, e.g., the Tully catalog (1988):
they suggest 10 Mpc instead of 2.4 Mpc which would correspondingly increase
the values of both $L_{\rm x}$ and $L_{\rm B}$.

\subsection {NGC\,1167}
The galaxy is a well-known radio source (4C\,+34.09) 
and has been intensively studied at radio wavelengths (e.g., Long et al. 1966,
Condon \& Dressel 1978, Bridle \& Fomalont 1978, Sanghera et al. 1995).   
Optical spectra were presented by, e.g., Wills (1967), Wills \& Wills (1976), and 
Gelderman \& Whittle (1994). 
Despite earlier suspicions, Ho et al. (1997; H97 hereafter) did not detect a broad component
in H$\alpha$. 
An upper limit for the X-ray luminosity
derived from \ein observations, $L_{\rm x}^{0.5-3.5 \rm keV} < 5\,10^{41}$ erg s$^{-1}$,
was reported by Dressel \& Wilson (1985; see also Canizares et al. 1987, Fabbiano et al. 1992). 

The source is undetected in the PSPC pointing, which might be partly traced back to the large 
\NH value in its direction, $N_{\rm gal} = 13.3\,10^{20}$ cm$^{-2}$. 
We estimate an upper limit for the countrate
of 0.005 cts/s, from the conservative assumption of a countrate less than that of the 
weakest detected source in the field of view at similar off-axis angle.

  \begin{figure*}[t]
\psfig{file=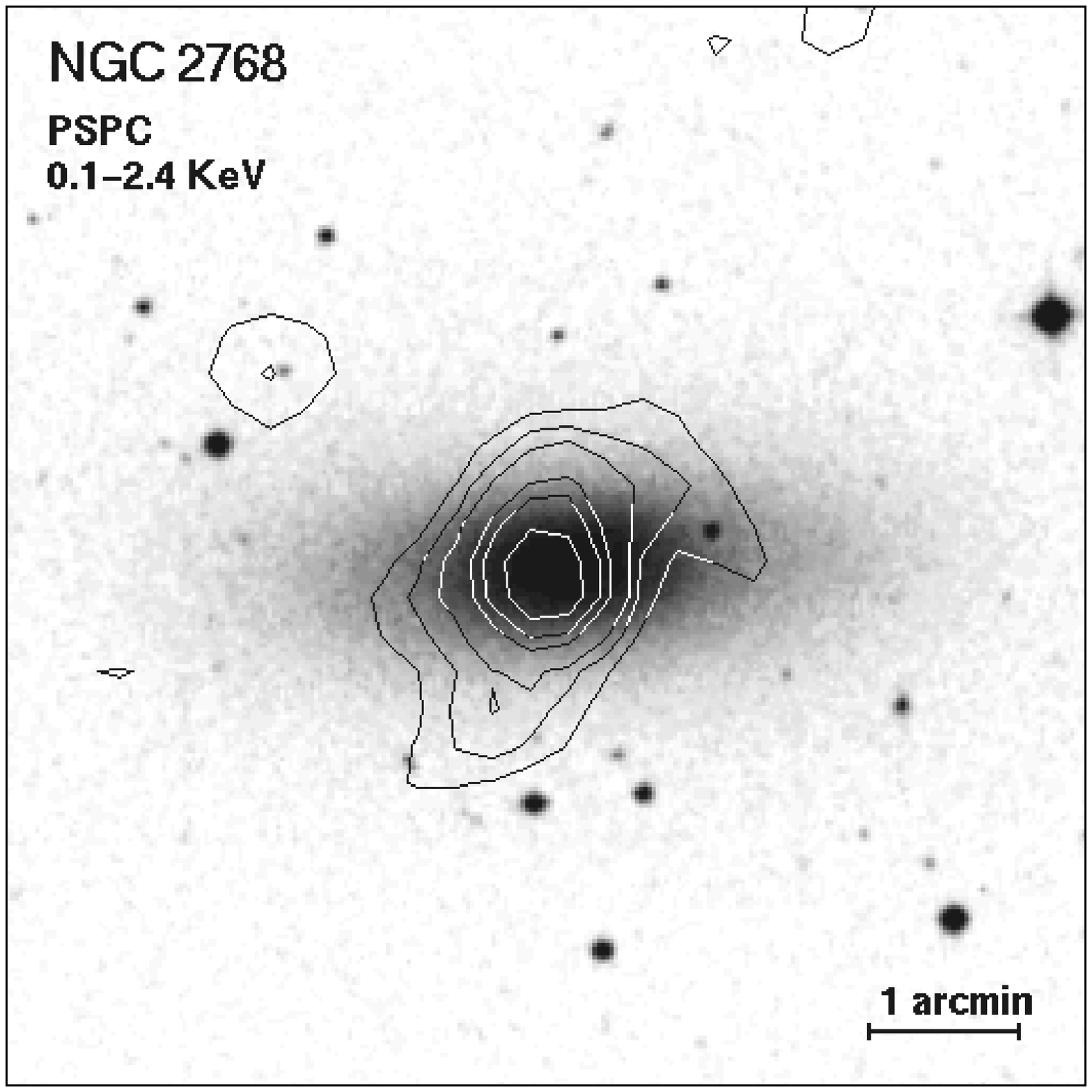,width=5.7cm}
\vspace*{-5.7cm} \hspace*{5.9cm}
\psfig{file=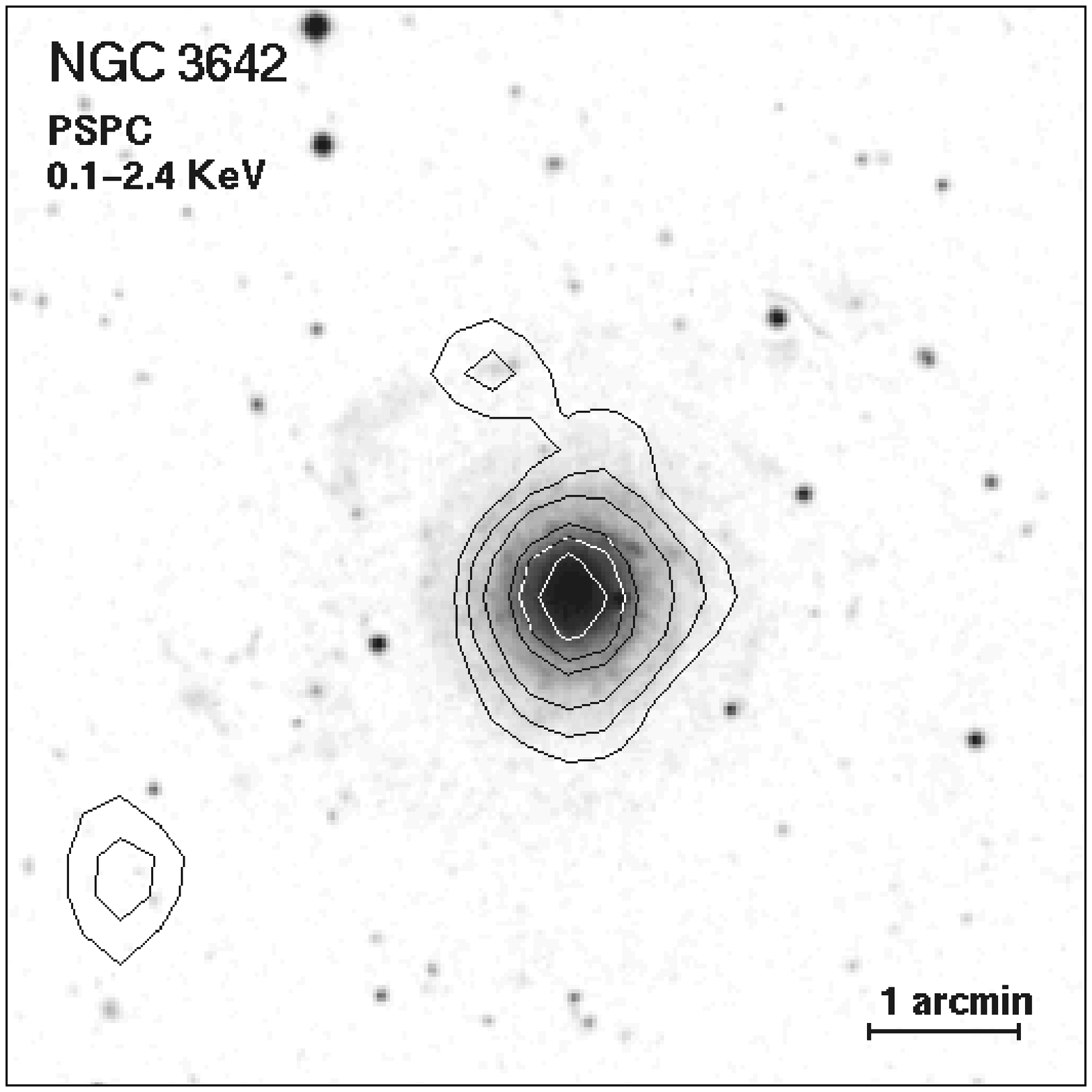,width=5.7cm}
\vspace*{-1.0cm}\hspace*{0.2cm}
\psfig{file=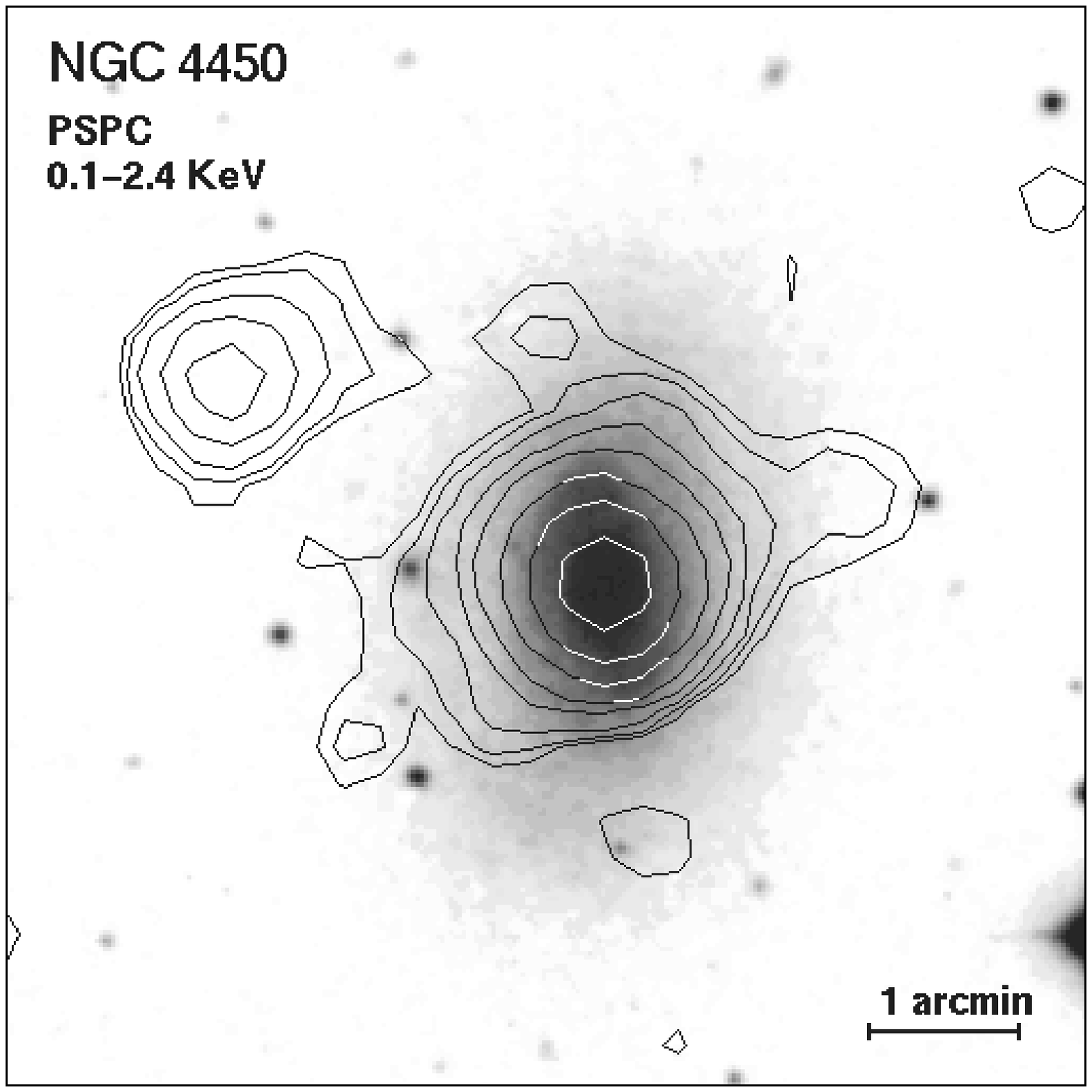,width=5.7cm}
\vspace*{1.3cm} 
 \caption[cont]{Contour plots for the X-ray emission of the three brightest on-axis
                PSPC sources overlaid on optical images from the digitized POSS.
NGC\,2768: contours are shown for 0.8, 1.1, 1.4, 2.1, 2.5, 3.1$\sigma$ above the
background; NGC\,3642: contours are 1.2, 1.8, 2.5, 3.9, 4.7, 5.8, 7.4$\sigma$;
NGC\,4450: contours are 1.2, 1.5, 1.9, 2.8, 4.3, 8.2, 15.9, 31$\sigma$ above the
background. 
}
\vspace*{0.5cm} 
\psfig{file=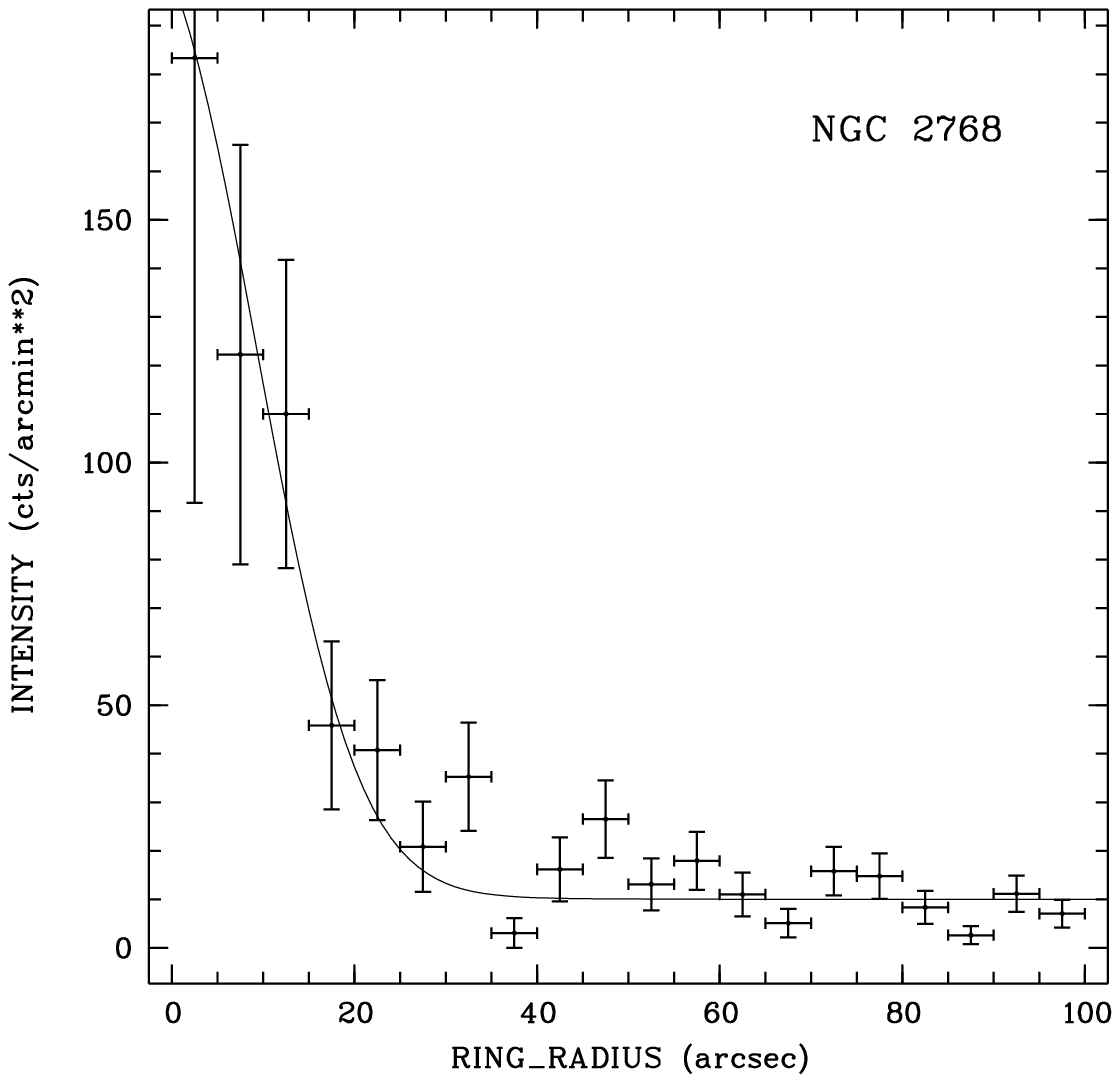,width=5.7cm}
\vspace*{-5.2cm} \hspace*{5.9cm}
\psfig{file=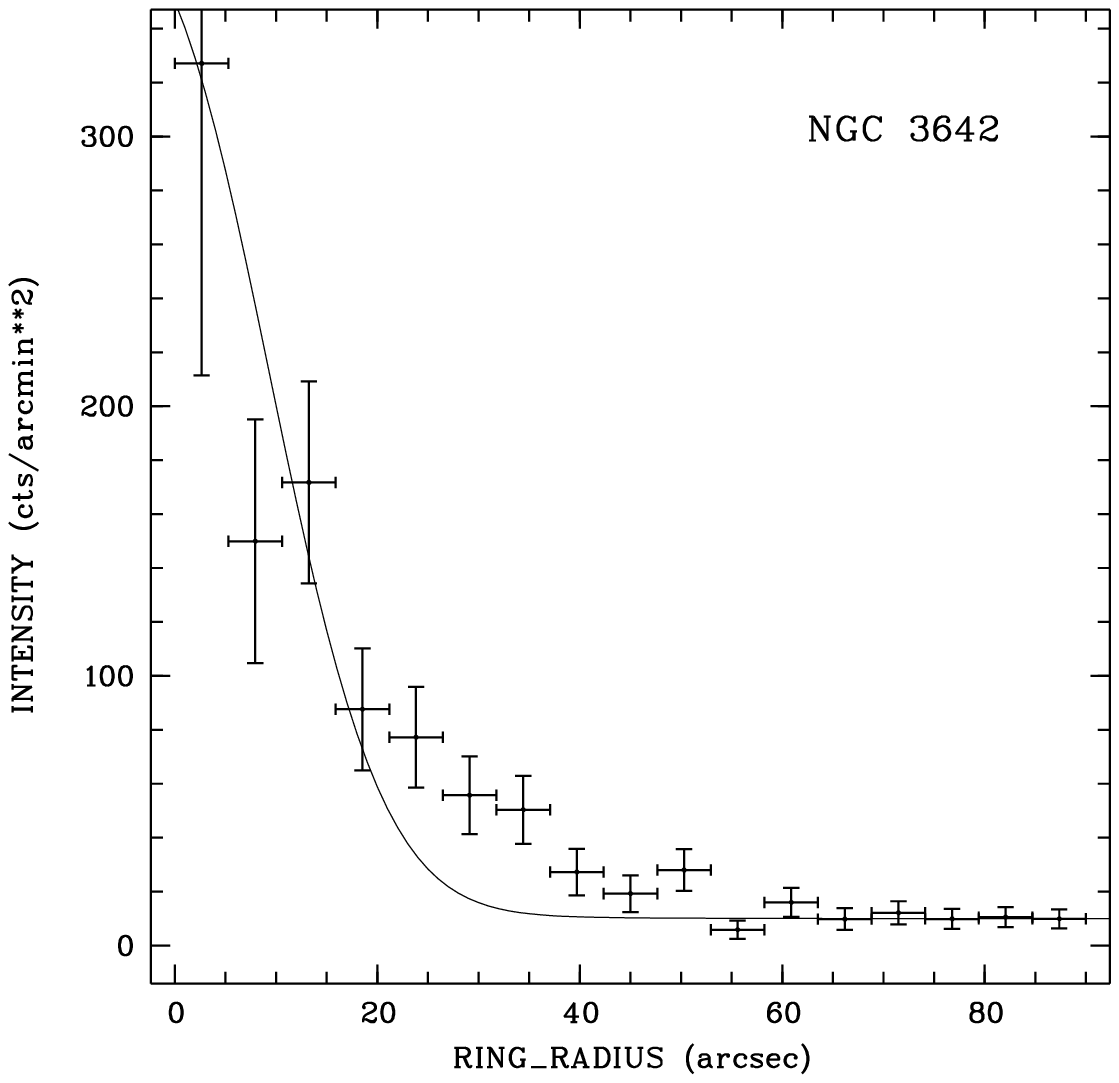,width=5.7cm}
\vspace*{-1.0cm}\hspace*{0.2cm}
\psfig{file=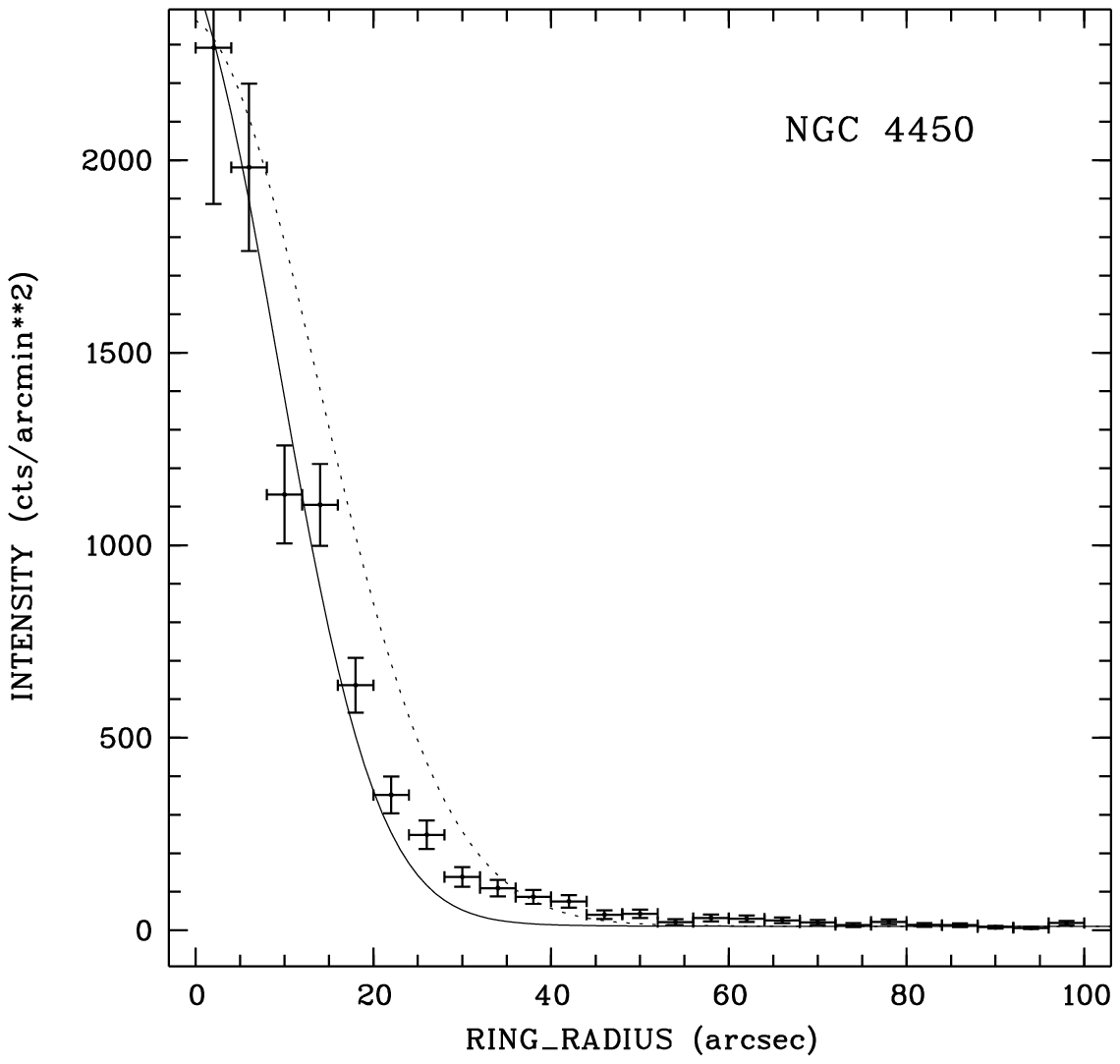,width=5.7cm}
\vspace*{1cm}
 \caption[radprof]{Azimuthally averaged radial profiles of the X-ray emission
from the three brightest on-axis PSPC sources compared to the PSPC's point spread function
for a point source (solid line: at 1 keV, dotted line: at 0.5 keV). The bulk of the X-ray emission is
consistent with arising from a point source.  }
\end{figure*}

\subsection {NGC\,2768}
No broad H$\alpha$ component
was detected by H97. Weak CO emission was found by Wiklind et al. (1995).
The source is included in a sample of galaxies by Davis \& White (1996) 
who fit a Raymond-Smith model and find $kT \approx 3$ keV for metal abundances 0.2 $\times$ solar, and
absorption of \NH = $1.9\,10^{20}$ cm$^{-2}$, less than the Galactic value. 

The source seems to be slightly variable from the first to the second pointing
with a drop in countrate from 0.021$\pm$0.002 cts s$^{-1}$ to 0.013$\pm$0.003 cts s$^{-1}$.
The short-term light curve (first pointing) shows constant source flux. 

Spectral fits were performed for the deeper PSPC observation only.
Neither a single powerlaw with $N_{\rm H}=N_{\rm gal}$ nor emission from a Raymond-Smith
plasma with solar abundances provides a successful X-ray spectral fit.
The fit becomes acceptable for very subsolar abundances.  
In that case we find a lower temperature than Davis \& White (1996)
(and cold absorption consistent with the Galactic value, which should not
be underpredicted). This $T$ also is more consistent with the $kT-\sigma$ relation
of Davis \& White. 

A comparison of the source's spatial extent with the theoretical
point spread function (PSF) of a point source shows that most of the
X-ray emission is consistent with arising from a point source (Fig. 3).
There may be some extended emission at weak levels (Fig. 2). A nearby weak 
second source is detected with a countrate of 0.003$\pm$0.001 cts/s.
It coincides with a stellar objects on a POSS plate. 

\subsection {NGC\,3642}

Some X-ray properties of this LINER were earlier examined by K95 who fit
a powerlaw model to the \ros PSPC spectrum and found the HRI source extent
to be consistent with a point source.
A broad component in H$\alpha$ is present (e.g., K95).
Barth et al. (1998) reported the detection of a compact nuclear UV source based
on HST data, and conclude that the extrapolation of the UV continuum, assuming
an AGN-like shape, would provide enough ionizing photons to power the NLR emission
of this galaxy.  

We neither detect short-time variability nor variability between the two
PSPC observations separated by 5 months. 

The spectrum is best fit by a Raymond-Smith model of heavily depleted
abundances, around 0.03 $\times$ solar  (or, alternatively, by a powerlaw
model with some excess absorption, confirming K95). 
 
A comparison with the PSF of the PSPC shows that most of the X-ray emission
arises from an unresolved source (Fig. 3). 
 There is a second source nearby with a countrate of 0.0025 $\pm$0.0007 cts/s.
Its position falls close to two star-like knots projected onto (or in)
one of the spiral arms of NGC\,3642 (they could be HII regions or foreground stars).
If the X-ray source is intrinsic to NGC\,3642, 
its luminosity of $L_{\rm x} = 1.8\,10^{39}$ erg s$^{-1}$ 
assuming a powerlaw spectral shape as described in Sect. 3
is fairly high. For instance, its exceeds the Eddington luminosity of a
solar mass black hole by a factor $\sim$10. One possible interpretation
is a powerful X-ray binary with either
a super-eddington low-mass black hole or a massive black hole.
We note in passing that no optical supernova was detected in NGC\,3642.

\subsection {NGC\,3898}
There have been several studies of this galaxy in the optical
(e.g., Burbidge \& Burbidge 1965, Barbon et al. 1978, Mizuno \& Hamajima 1986, and references
given in van Driel \& van Woerden 1994). 
H97 tentatively concluded that broad H$\alpha$ is absent
from the optical spectrum. 
21\,cm HI observations with the WRST were presented by van Driel \& van Woerden (1994). 

The source is quite weak with only about 50 detected photons, very close to the limits
of meaningful $\chi^2$ spectral fits. Therefore, we only applied a powerlaw model
with fixed Galactic absorption. This results in \G=--2.1 and gives an acceptable fit.

   \begin{table*}     
     \caption{ Results of spectral fits; first columns: powerlaw with $N_{\rm H}=N_{\rm gal}$
               (for results with $N_{\rm H}$ as free parameter see text), last columns 
               Raymond-Smith model. Fluxes (absorption corrected)
               and luminosities refer to the (0.1-2.4 keV) energy band.
               For the sources
               that did not allow a PSPC spectral analysis, we assumed a powerlaw of
               photon index \G=--1.9 
               plus an amount of cold absorption corresponding to the Galactic value in direction
               of the individual galaxies (Dickey \& Lockman 1990) to
               calculate fluxes and luminosities (if RASS and pointed observations were
               available, we used the observation with the longest exposure time to derive
               $f$ and $L$). For some of the sources only observed
               during the RASS, only upper limits on count rates are available. 
               The errors in \G and kT, given for successful spectral fits only,
               are quoted at the 68.3\% confidence level. 
               Metal abundances reported in the last but one column were fixed. 
        }
     \label{fitres}
      \begin{tabular}{llccccccc}
      \noalign{\smallskip}
      \hline
      \noalign{\smallskip}
        object & $N_{\rm gal}$ & $\Gamma_{\rm x}$ & $f_{\rm pl}$ & $\log L_{\rm pl}$ & $\chi^2_{\rm red}$ &
                               $kT_{\rm rs}$  & $Z$   
                               & $\chi^2_{\rm red}$ \\
       \noalign{\smallskip}
      \noalign{\smallskip}
           & 10$^{21}$ cm$^{-2}$ & & 10$^{-12}\,$erg\,cm$^{-2}$\,s$^{-1}$ & erg\,s$^{-1}$ & &
                                    keV & sol & \\  
       \noalign{\smallskip}
      \hline
      \hline
      \noalign{\smallskip}
      \noalign{\smallskip}
  NGC\,404 & 0.509 &--1.9$^1$ & 0.07 & 37.7 & & & & \\  
\noalign{\smallskip}
  NGC\,1167 & 1.33 &--1.9$^1$ & $<$0.18~~ & $<$40.95~~~ & & & &  \\
\noalign{\smallskip}
  NGC\,2768 & 0.383 & --2.2$\pm{0.3}$ & 0.58 & 40.6 & 3.1 & 3.7 & 1.0 & 5.3 \\  
            &       &       & & &     & 0.5$\pm{0.1}$ & 0.1 & 0.8 \\
\noalign{\smallskip}
  NGC\,3642 & 0.073 & --1.8$\pm{0.2}$ & 0.18 & 40.2 & 1.4 & 5.5 & 1.0 & 3.6 \\
            &       &       & & &     & 1.1$\pm{0.3}$ & 0.1 & 1.5 \\
            &       &       & & &     & 0.7$\pm{0.3}$ & 0.05 & 0.8 \\
\noalign{\smallskip}
  NGC\,3898 & 0.106 & --2.1$\pm{0.5}$ & 0.29 & 40.2 & 0.8 & & & \\
\noalign{\smallskip}
  NGC\,4036 & 0.192 &--1.9$^1$ & 0.20 & 40.2& & & & \\
\noalign{\smallskip}
  NGC\,4419 & 0.255 &--1.9$^1$ & $<$2.43~~ & $<$40.9~~~& & & & \\
\noalign{\smallskip}
  NGC\,4450 & 0.245 & --2.0$\pm{0.1}$ & 1.89 & 40.8 & 1.1 & 3.6 & 1.0 & 10 \\
            &       &       &      & &     & 1.4 & 0.1 & 3.9 \\
\noalign{\smallskip}
  NGC\,5371 & 0.110 & --1.6$\pm{0.1}$ & 0.40 & 40.8 & 1.1 & 5.3 & 1.0 & 5.1 \\
            &       &       &      & &     & 1.4 & 0.1 & 3.2 \\
            &       &       &      & &     & 0.8$\pm{0.2}$ & 0.01 & 1.8 \\
\noalign{\smallskip}
  NGC\,5675 & 0.100 &--1.9$^1$ & $<$0.18~~ & $<$40.8~~~& & & & \\
\noalign{\smallskip}
  NGC\,5851 & 0.258 &--1.9$^1$ & $<$1.40~~ & $<$42.1~~~& & & & \\
\noalign{\smallskip}
  NGC\,6500 & 0.753 &--1.9$^1$& 0.19 & 40.6 & & & & \\
\noalign{\smallskip}
  IC\,1481  & 0.566 &--1.9$^1$ & $<$0.49~~ & $<$41.6~~~& & & & \\
\noalign{\smallskip}
      \noalign{\smallskip}
      \hline
      \noalign{\smallskip}
  \end{tabular}

\noindent{\small $^{(1)}$ fixed
}
   \end{table*}

\subsection {NGC\,4450}
A fairly weak broad H$\alpha$ line is probably present in the
optical spectrum (Staufer 1982,
H97). For an optical image see, e.g., Sandage (1961). The HII region population
of the galaxy was studied by Gonzalez Delgado et al. (1997). 
An \ein IPC image is shown in Fabbiano et al. (1992). They derive an 
(0.2-4 keV) X-ray flux $f_{\rm x} = 11.5\,10^{-13}$ erg cm$^{-2}$ s$^{-1}$
under the assumption of a thermal bremsstrahlung spectrum with $kT$=5\,keV.   

Again, we do not detect short-timescale variability. 

The source is quite bright and nearly 2000 photons are available for
the spectral analysis (we used the deepest pointing). 
No Raymond-Smith fit is possible. When allowing \NH to be free, it underpredicts
the Galactic value. If subsolar abundances are allowed,
the best fit requires abundances
less than 1/100 solar and that fit is still unsatisfactory.
In contrast, a single powerlaw with \G = --2.0 near the AGN-canonical value
(e.g., Pounds et al. 1994; Svensson et al. 1994)  
gives an excellent fit. If \NH is treated as free parameter, the Galactic value is
recovered. We derive a soft X-ray luminosity of $L_{\rm x} = 10^{40.8}$ erg s$^{-1}$,
the highest value found among the present galaxies.
The corresponding (0.5-4.5 keV) X-ray luminosity is $L_{\rm x}^{0.5-4.5 \rm keV} = 10^{40.6}$ erg s$^{-1}$,
a factor $\sim$5 above the value expected from the stellar contribution 
using the relation of Canizares et al. (1987). 
We note that Tully's catalog places NGC\,4450 at the distance of the Virgo cluster.
If the galaxy is instead located in the sheet of galaxies behind the Virgo
cluster, the luminosities inferred above increase correspondingly.   

A comparison with the PSF of the PSPC shows that most of the X-ray emission
is consistent with arising from a point source. At weak emission levels 
there is evidence for source extent (Fig. 3; several of the structures 
are seen in both, the soft (0.1-0.5 keV) and hard (0.5-2.4 keV) band). 
Again, there is a nearby second source. Its countrate is 
0.011 $\pm$0.002 cts/s and since the pointing is deep, a spectral
analysis is possible. A powerlaw spectral fit gives a 
spectrum similar to NGC\,4450 itself,
but a bit softer with \G=--2.4. At the distance of NGC\,4450 
this corresponds to a luminosity of $L_{\rm x} = 7\,10^{39}$ erg s$^{-1}$.
The source is also present in the second PSPC pointing of slightly lower
exposure time (Table 1). Its countrate is constant. 
Inspection of the POSS plates does not reveal any optical counterpart.
Neither is there any X-ray source visible in the \ein IPC image (see Fig. 7
of Fabbiano et al. 1992).   
The `reality' of these nearby sources is examined in Section 4.9.

\subsection {NGC\,5371}
The galaxy was classified as a LINER by Rush et al. (1993).
Elfhag et al. (1996) presented CO measurements and 
suggested NGC\,5371 to be a good candidate for a post-starburst galaxy (Koorneef 1993).
The rotation curve was measured by, e.g., Zasov \& Sil'chenko (1987).
Gonzalez Delgado et al. (1997) studied the HII region population. 

The X-ray lightcurve does not show short-timescale variability (the 
countrate in individual bins
falls slightly outside the 1$\sigma$ error, occasionally, but this is most likely
due to the closeness of the source to the PSPC support grid structure).  

The source appears to be extended or double. Thus, photons from the
total emission region were first extracted for analysis. 
A spectrum was fit to this `double' source (since their
contributions cannot be disentangled from each other safely;
it is these results that are listed in Table 2).
In that case, a powerlaw of index \G $\simeq -1.6$ yields a successful
X-ray spectral fit. The cold absorption, if treated as free parameter,
underpredicts the Galactic value, and even more so if a RS model is
applied. 
The latter type of model only provides an acceptable fit if the 
metal abundances are
depleted below 0.01 $\times$ solar. 

Secondly, since the optical position of NGC\,5371 falls on the northernmost of
the two sources, source photons centered on the optical position of the galaxy
were extracted within a circular region of diameter 250\arcsec. In this case, the X-ray
spectrum is dominated by the northernmost source, but the second one contributes
to some extent.
The spectral analysis then yields a best fit in terms of a powerlaw with
\G $\simeq -1.96$  ($\chi^2_{\rm red}$=0.5), and $N_{\rm H}$ recovers the Galactic value if
treated as free parameter. Again, RS emission can only successfully describe the 
spectrum for heavily depleted abundances.   

The source appears double, or extended. No HRI observation is available for a
more detailed study of the spatial extent.

\subsection {NGC\,6500}
There are several lines of evidence for the presence of a nuclear
 outflow or wind as judged from  radio continuum emission 
measurements (Unger et al. 1989) and  optical emission lines (Gonzalez Delgado \& 
 Perez 1996). H97 did not detect broad  H$\alpha$. 
 Barth et al. (1997) using HST data concluded that the resolved UV emission
 of NGC\,6500 is likely dominated by massive stars. They also 
 derived a \ros HRI flux for this galaxy, 
 assuming a spectrum with \G=--2.   

Although NGC\,6500 is detected in the HRI observation, the low number of source photons prevents
a more detailed analysis in terms of source extent or variability.

\subsection { Origin of nearby sources}
In PSPC observations of several galaxies (NGC\,2768, NGC\,3642, NGC\,4450) 
there is a second
source detected 
near the target source with a countrate always roughly 1/10
of the central source. The same was found for NGC\,4736 by
Cui et al. (1997) who considered the second source to be real
and of transient nature 
due to its presence in the PSPC and absence in the
HRI observation. 
Given the similar locations relative to the central source,
and same (factor $\approx$ 1/10) relative countrates, 
we suspected these second sources 
to be an instrumental artifact, namely ghost imaging (Briel et al. 1994) to
be at work. 

To more closely examine this problem we selected the second source near NGC\,4450
since it is the brightest, thus allowing the most detailed analysis, and since 
no potential optical counterpart shows up near the X-ray position. 
We extracted the photons around the X-ray center of the source
and made several tests. However, we find no indications of ghost imaging:
Source photons do not exclusively cover the very soft channels
(ghost imaging only operates below $\sim$ 0.2 keV; Nousek \& Lesser 1993, Briel et al. 1994), 
and the source
is not fixed in detector coordinates but follows the wobble. 

Using the X-ray $\log N - \log S$ distribution of Hasinger et al. (1994), we expect only
0.17 sources of X-ray flux greater or equal to that of the source near NGC\,4450
in a region of size 10\arcmin~$\times$~10\arcmin.
For a discussion of an excess of bright X-ray sources around nearby galaxies
see Arp (1997, and references therein).  

Unfortunately, no HRI observation of NGC 4450 is available for further
scrutiny. It will certainly be interesting to check for the presence
of the second source once further high spatial resolution observations of NGC\,4450
become available.

\section{Discussion}

\subsection {X-ray luminosity and $L_{\rm x}$/$L_{\rm B}$ ratio}

The objects examined span a luminosity range from
$L_{\rm x} \simeq 10^{37.7} - 10^{40.8}$ erg s$^{-1}$ in the
(0.1-2.4 keV) band.
There is still some bias towards selecting the high $L_{\rm x}$ LINERs.
This does not hold for the \ros survey data but given the short exposure
times of typically 400s upper limits, although already meaningful, are not
very restrictive concerning the low-luminosity end.

None exceeds the limit of $L_{\rm x} \approx 10^{42}$ erg\,s$^{-1}$ which
is usually taken as indicative for the presence of a `normal' AGN
(e.g., Wisotzki \& Bade 1997).
Also, none reaches the high $L_{\rm x}$ usually observed 
for ellipticals in the group/cluster environment (e.g., Brown \& Bregman 1998,
Irwin \& Sarazin 1998, Beuing et al. 1999)
which occasionally show LINER-like emission lines.

Most of the present objects fall in the intermediate 
$L_{\rm x}$/$L_{\rm B}$ range (Fig. 4). 
Since the majority of LINERs are found in bulge-dominated 
early-type galaxies (e.g., Ho 1998) 
the same emission mechanisms might contribute to the observed X-ray luminosity.
Among the suggestions for early-type galaxies  
are accumulated stellar winds, SN heating and cooling flows 
(see Pellegrini 1999 for a recent overview).

The low $L_{\rm X}/L_{\rm B}$ systems
are dominated by discrete sources, mainly LMXBs (e.g., Canizares et al. 1987,
Irwin \& Sarazin 1998, Irwin \& Bregman 1999). 
Among the present sample, this holds best for NGC\,404.

Further clues on the emission mechanism can be drawn from the 
observed spectral shapes. 

  \begin{figure}[t]
\hspace*{0.2cm}
\psfig{file=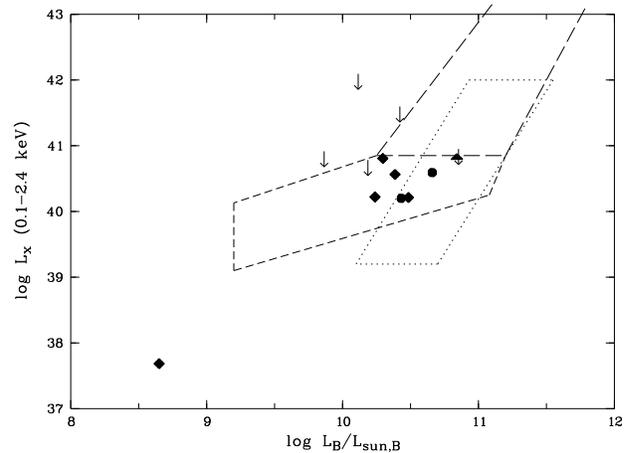,width=8.3cm}
 \caption[lxlb]{LINERs in the $L_{\rm x}-L_{\rm B}$ diagram.
Filled symbols denote detected sources of different morphological
type (circles: S0s, lozenges: SAs, triangles: SABs),
arrows mark upper limits.
For comparison, the regions populated by some samples of elliptical galaxies
are marked (dotted line: Brown \& Bregman 1998, long-dashed: Beuing et al. 1999 (detections),
short-dashed: Beuing et al. 1999 (upper limits)).
The X-ray emission from NGC\,404 (lower left)
is consistent with originating completely from discrete stellar sources.
  }
\label{lxlb}
\end{figure}

\subsection{X-ray spectral shapes} 

The LINERs analyzed here show some spectral variety. 
Some of them are best described by a powerlaw model
of photon index similar to that observed in 
AGN (e.g., Schartel et al. 1996a,b, 1997), the others are best
fit by Raymond-Smith (RS) emission of a very sub-solar abundances plasma (see Table 2). 

We consider the RS model with heavily depleted gas-phase
metal abundances to be quite unrealistic, and 
as previously found for other galaxies (ellipticals, AGN; e.g., 
Matsushita et al. 1997, 
Buote \& Fabian 1998, Komossa \& Schulz 1998) we prefer the alternative of a two-component spectrum,
consisting of contributions from both, a powerlaw (or other hard component) and thermal RS emission
of $\approx$solar abundances gas.
Such models were not fit, though, due to the quite low number of photons
available per spectrum.
Other possibilities (than the two-component solution) include a contribution
from Fe\,L emission, not yet fully modeled (but see Buote \& Fabian 1998
who excluded the FeL region from fitting ASCA spectra of early type galaxies,
and still find similar results concerning abundances), a multi-temperature distribution
of the emitting medium (e.g., Strickland \& Stevens 1998) and  
the possibility that the X-ray emitting gas is far out of collisional-ionization equilibrium 
(e.g., Breit\-schwerdt \& Schmutzler 1994, 1999, Komossa et al. 1999; see B{\"o}hringer 1998
for a recent review).  

In case of the two-component interpretation -- as indeed observed by \asca 
with its broader X-ray energy range for several early-type galaxies
(e.g., Matsushita et al. 1994, Matsumoto et al. 1997, Buote \& Fabian 1998)
and some LINERs (e.g., Ptak et al. 1999) -- the hard  
component could be due to stellar sources, namely LMXBs, 
or a low-luminosity AGN, the soft component 
due to emission from hot gas (see previous Section).  

In the case of NGC\,6500 the X-ray emission might be related to the outflow/wind for which
optical and radio evidence
was reported (Unger et al. 1989, Gonzales\,Delgado \& Perez 1996), 
in analogy to the X-ray emission associated
with starburst-driven winds observed in
several starburst galaxies (e.g., Heckman et al. 1990, Schulz et al. 1998).
It is also interesting to note that in two further cases, NGC\,4450 and NGC\,2768,
the weak extended emission appears roughly perpendicular to the galaxy's disk and could
result from an outflow. 

We interpret the powerlaw spectral component, in those cases where it
dominates the spectrum, as arising most likely from 
a low-luminosity active nucleus, since the inferred
luminosities are above those  
expected from discrete stellar sources and the 
powerlaw indices derived are in the range \G $\simeq -1.6$ to $-2.1$ (Table 2)
similiar to what is observed for AGNs. 
However, we cannot exclude a more complex situation      
where the superposition of several different emission components 
mimics a single AGN-like powerlaw.

For a detailed spatial and spectroscopic disentanglement of the contributing components (thermal
RS-like emission, presumably extended, and a pointlike powerlaw-like component)
and a better determination of the metal abundances
high-spectral resolution observations with
future X-ray observatories like XMM, AXAF or Spectrum-X-Gamma will be very useful.

\subsection{X-ray variability}
We do not find evidence for X-ray variability on the timescale of hours/days.
This also holds for the LINERs examined by Ptak et al. (1998) and some
further sources, and  
is not in line with a continuation of the trend of higher variability at
lower luminosity seen in AGN. 
It {\em is} consistent with the  
presence of advection-dominated accretion disks (e.g., Abramowicz et al. 1988, 
Narayan \& Yi 1994) in
LINERs as suggested by Ptak et al. (1998; see also Lasota et al. 1996).

\section {Summary and conclusions}

We have investigated the soft X-ray properties of a number of LINERs
based on \ros survey and pointed PSPC and HRI observations.   

Luminosities range between $\log L_{\rm x} = 37.7$ (NGC\,404) 
and 40.8 (NGC\,4450). The ratios $L_{\rm x}$/$L_{\rm B}$, when compared to
early-type galaxies, are located in the intermediate region 
and similar emission mechanisms may contribute to the observed X-ray luminosities.

Whereas the bulk of the X-ray emission is consistent with arising from a point source
there is some extended emission seen at weak emission levels in some sources.

Spectra are best described by either a powerlaw of photon index \G $\approx -2$  
(NGC\,3898, NGC\,4450, NGC\,5371), or a Raymond-Smith model
with heavily depleted abundances (NGC\,2768, NGC\,3642).
Since the inferred metal abundances are implausibly low 
we take 
this latter model as indication of a more complex
soft X-ray spectral shape/ emission mechanism 
(e.g., a powerlaw or other hard component
 plus RS emission from gas with $\approx$solar
abundances; or plasma out of equilibrium). 
Some sources are best described by a single AGN-like powerlaw 
with X-ray luminosity above that expected from discrete stellar
sources. These spectra most likely indicate the presence of
low-luminosity AGNs in the centers of the LINER galaxies.   

The absence of (short-timescale) variability is consistent 
with    
the earlier suggestion that LINERs may
accrete in the advection-dominated mode. 
Only one source (NGC\,2768) seems to be slightly variable on a timescale of months. 

For several LINER galaxies nearby second X-ray sources are discovered with 
countrates roughly one-tenth those of the target sources.  
We cannot identify an obvious detector effect. If at the distances
of the galaxies, their luminosities are on the order of several $10^{39}$ erg s$^{-1}$. 

Given the spectral variety of LINERs with contributions from several 
emission components, future studies of both, spectra of individual objects 
as well as larger samples will certainly give further insight into
the LINER phenomenon which provides an important link between active and `normal' galaxies.

\begin{acknowledgements}
H.B. and St.K. thank the Verbundforschung for support under grant No. 50\,OR\,93065.
J.P.H. acknowledges support from the Smithsonian
Institution and NASA grant NAGW-201.
We thank Andreas Vogler for providing the software to plot the overlay contours of 
Fig. 2. 
The \ros project has been supported by the German Bundes\-mini\-ste\-rium
f\"ur Bildung, Wissenschaft, Forschung und Technologie 
(BMBF/DLR) and the Max-Planck-Society.
This research has made use of the NASA/IPAC extragalactic database (NED)
which is operated by the Jet Propulsion Laboratory, Caltech,
under contract with the National Aeronautics and Space
Administration.
The optical images shown are based on photographic data of the
National Geographic Society -- Palomar
Observatory Sky Survey (NGS-POSS) obtained using the Oschin telescope
on Palomar Mountain.

\end{acknowledgements}


\begin{thebibliography}{}

\bibitem{}  Abramowicz M.A., Czerny B., Lasota J.-P., Szuszkiewicz E., 1988, ApJ 332, 646 

\bibitem{} Anders E., Grevesse N., 1989, Geochimica \& Cosmochimica Acta 53, 197

\bibitem{} Arp H., 1997, A\&A 319, 33 

\bibitem{} Baars J.W.M., Wendker H.J., 1976, A\&A 48, 405

\bibitem{} Barbon R., Benacchio L., Capaccioli M., 1978, A\&A 65, 165  

\bibitem{} Barth A.J., Reichert G.A., Ho L.C., et al., 1997,
	    AJ 114, 2313

\bibitem{} Barth A.J., Ho L.C., Filippenko A.V., Sargent W.L.W., 1998, ApJ 496, 133

\bibitem{} Beuing J., D\"obereiner S., B\"ohringer H., Bender R., 1999,
            MNRAS 302, 291 


\bibitem{}Binette L., 1984, in `The Messenger' 38, 13

\bibitem{}Binette L., 1985, A\&A 143, 334 

\bibitem{}Binette L., 1986, in `Structure and evolution of active galactic nuclei',
    G. Giuricin et al. (eds), 475, Reidel: Dordrecht 

\bibitem{} B\"ohringer H., 1998, in `The local bubble and beyond', D. Breitschwerdt,
           M.J. Freyberg, J. Tr\"umper (eds), LNP 506, 341 

\bibitem{} Bohlin R.C., Savage B.D., Drake J.F., 1978, ApJ 224, 132

\bibitem{} Breitschwerdt D., Schmutzler T., 1994, Nat 371, 774 

\bibitem{} Breitschwerdt D., Schmutzler T., 1999, to appear in A\&A; astro-ph/9902268

\bibitem{} Bridle A.H., Fomalont E.B., 1978, MNRAS 185, 67p

\bibitem{}Briel U., Aschenbach B., Hasinger G. et al., 1994, ROSAT
user's handbook, MPE: Garching

\bibitem{} Brown B.A., Bregman J.N., 1998, ApJ 495, L75

\bibitem{} Buote D.A., Fabian A.C., 1998, MNRAS 296, 977

\bibitem{} Burbidge E.M., Burbidge G.R., 1962, ApJ 135, 694

\bibitem{} Burbidge E.M., Burbidge G.R., 1965, ApJ 142, 634 

\bibitem{} Canizares C.R., Fabbiano G., Trinchieri G., 1987, ApJ 312, 503

\bibitem{} Cui W., Feldkun D., Braun R., 1997, ApJ 477, 693 

\bibitem{} Condon J.J., Dressel L., 1978, ApJ 221, 456

\bibitem{} Contini M., 1997, A\&A 323, 71 

\bibitem{} Davis D.S., White III R.E., 1996, 470, L35 

\bibitem{} de Vaucouleurs G., de Vaucouleurs  A., Corwin J.R., et al., 1991, 
            Third reference catalogue of
            bright galaxies, New York: Springer

\bibitem{} Dickey J.M., Lockman F.J., 1990, ARA\&A, 28, 215

\bibitem{} Dopita M., Allen M., Bicknell G.V., et al., 1996, in `The physics
   of LINERs in view of recent observations', M. Eracleous et al. (eds),
   ASP conf ser. 103, 44 

\bibitem{} Dressel L.L., Wilson A.S., 1985, ApJ 291, 668 

\bibitem{} Elfhag T., Booth R.S., H{\"o}glund B., Johansson L.E.B., Sandqvist Aa., 1996, 
            A\&AS 115, 439 

\bibitem{} Ehle M., Pietsch W., Beck R., 1995, A\&A 295, 289

\bibitem{} Eracleous M., Livio M., Binette L., 1995, ApJ 445, L1

\bibitem{} Fabbiano G., Kim D.-W., Trinchieri G., 1992, ApJS 80, 531

\bibitem{} Falcke H., 1998, Rev. mod. Astron. 11, 245 

\bibitem{} Falcke H., Wilson A.S., Ho L.C., 1997, in proc. of `Relativistic Jets in AGN',
           p. 13   

\bibitem{} Filippenko A.V., 1989, in `Active galactic nuclei',
	    D.E. Osterbrock and J.S. Miller, (Dordrecht: Kluwer), 495

\bibitem{} Filippenko A.V., 1993, in `The nearest active galaxies', J. Beckman (ed.),
           Ap\&SS 205 

\bibitem{} Filippenko A.V., Sargent W.L.W., 1985, ApJS 57, 503    

\bibitem{} Ferland G., Netzer H., 1983, ApJ 264, 105

\bibitem{} Fosbury R.A.E., Mebold U., Goss W.M., Dopita M.A., 1978, MNRAS 183, 549

\bibitem{} Gelderman R., Whittle M., 1994, ApJS 91, 491

\bibitem{} Gonzales Delgado R., Perez E., 1996, MNRAS 281, 1105

\bibitem{} Gonzales Delgado R., Perez E., Tadhunter C., Vilchez J.M., Rodriguez-Espinosa J.M.,
            1997, ApJS 108, 155

\bibitem{} Halpern J., Steiner J.E., 1983, ApJ 269, L3

\bibitem{} Hasinger G., Burg R., Giacconi R., et al., 1994, A\&A 275, 1

\bibitem{} Heckman T., 1980, A\&A 87, 152

\bibitem{} Heckman T., Armus L., Miley G.K., 1990, ApJS 74, 833 

\bibitem{} Ho L.C., 1998, to appear in `The 32nd COSPAR Meeting', Adv. Space Res.;
            astro-ph/9807273

\bibitem{} Ho L.C., Filippenko A.V., Sargent W.L.C., 1993, ApJ 417, 63 

\bibitem{} Ho L.C., Filippenko A.V., Sargent W.L.C., 1995, ApJS 98, 477

\bibitem{} Ho L.C., Filippenko A.V., Sargent W.L.C., Peng C.Y., 1997, ApJS 112, 391

\bibitem{} Huchra J., Burg R., 1992, ApJ 393, 90  

\bibitem{} Irwin J.A., Sarazin C.L., 1998, ApJ 499, 650

\bibitem{} Irwin J.A., Bregman J.N., 1999, ApJ 510, L21

\bibitem{} Keel W.C., 1983, ApJ 269, 466    

\bibitem{} Komossa S., Breitschwerdt D., B\"ohringer H., Meerschweinchen J., 1999,
           in `Astrophysical Dynamics, D. Berry et al. (eds), Ap\&SS, in press  

\bibitem{} Komossa S., Schulz H., 1998, A\&A 339, 345 

\bibitem{} Koorneef J., 1993, ApJ 403, 581

\bibitem{} Koratkar A., Deustua S.E., Heckman T., et al., 1995, ApJ 440, 132 

\bibitem{} Larkin J.E., Armus L., Knop R.A., Soifer B.T., Matthews K.,
           1998, ApJS 114, 59  

\bibitem{} Lasota J.P., Abramowicz M.A., Chen X., et al., 1996, ApJ 462, 142 

\bibitem{} Long R.J, Smith M.A., Stewart P., Williams P.J.S., 1966, MNRAS 134, 371

\bibitem{} Maoz D., Filippenko A.V., Ho L.C., et al., 1995, ApJ 440, 91

\bibitem{} Maoz D., Koratkar A., Shields J.C., et al., 1998, AJ 116, 55

\bibitem{} Matsumoto H., Koyama K., Awaki H., et al., 1997, ApJ 482, 133

\bibitem{} Matsushita K., Makishima K., Awaki H., et al., 1994, ApJ 436, L41

\bibitem{} Matsushita K., Makishima K., Rokutanda E., Yamasaki N., Ohashi T., 1997, ApJ 488, L125 

\bibitem{} Mizuno T., Hamajima K., 1986, PASJ 39, 211  

\bibitem{} Mould J., et al., 2000, ApJ, in prep. 

\bibitem{} Mushotzky R., 1982, ApJ 256, 92 

\bibitem{} Narayan R., Yi I., 1994, ApJ 428, L13   

\bibitem{} Nousek J., Lesser A., 1993, \ros Newsletter 8, 13 

\bibitem{} Pellegrini S., 1999, to appear in A\&A; astro-ph/9812325 

\bibitem{} Pfeffermann E., Briel U.G., Hippmann H., et al.,  
            1987, SPIE 733, 519

\bibitem{} Pietsch W., Trinchieri G., Vogler A., 1998, A\&A 340, 351  

\bibitem{} Pounds K.A., Nandra K., Fink H., Makino F., 1994, MNRAS 267, 193

\bibitem{} Predehl P., Schmitt J.H.M.M., 1995, A\&A 293, 889

\bibitem{} Ptak A., Yaqoob T., Mushotzky R., Serlemitsos P., Griffiths R.,
           1998, ApJ 501, L37

\bibitem{} Ptak A., Serlemitsos P., Yaqoob T., Mushotzky R., 
            1999, to appear in ApJS; astro-ph/9808159 

\bibitem{}  Raymond J.C., Smith B.W., 1977, ApJS 35, 419 

\bibitem{}  Roberts T.P., Warwick R.S., Ohashi T., 1999, MNRAS 304, 52

\bibitem{} Rush B., Malkan M.A., Spinoglio L., 1993, ApJS 89, 1  

\bibitem{} Sandage A., 1961, The Hubble Atlas of Galaxies, Wahington D.C.:
Carnegie Institution of Washington

\bibitem{} Sanghera H.S., Saikia D.J., Ludke E., et al., 1995, A\&A 295, 629 

\bibitem{} Schartel N., Walter R., Fink H., Tr\"umper J., 1996a, A\&A 307, 33

\bibitem{} Schartel N., Green P.J., Anderson S.F., et al., 1996b, MNRAS 283, 1015

\bibitem{} Schartel N., Schmidt M., Fink H., Hasinger G., Tr\"umper J., 1997, A\&A 320, 696

\bibitem{} Schulz H., Fritsch C., 1994, A\&A 291, 713 

\bibitem{} Schulz H., Komossa S., Bergh\"ofer T., Boer B., 1998, A\&A 330, 823

\bibitem{} Serlemitsos P., Ptak A., Yaqoob T., 1997, 
	   in `The physics of LINERs in view of recent observations',
	M. Eracleous et al. (eds), ASP conf. ser. 103, 70

\bibitem{} Shields J.C., 1992, ApJ 339, L27 

\bibitem{} Stauffer J. R., 1982, ApJ 262, 66 

\bibitem{} Stockdale C.J., Romanishin W., Cowan J.J., 1998, ApJ 508, L33

\bibitem{} Strickland D.K., Stevens I.R., 1998, MNRAS 297, 747 

\bibitem{} Stromberg G., 1925, ApJ 61, 353  

\bibitem{} Svensson R., 1994, ApJS 92, 585

\bibitem{} Terashima Y., Kunieda H., Misaki K., et al., 1998, ApJ 503, 212

\bibitem{} Tully R.B., 1988, Nearby galaxies catalog, 
           Cambridge University Press: Cambridge

\bibitem{} Tully R.B., Shaya E.J., 1984, ApJ 281, 31 

\bibitem{} Tr\"umper J., 1983, Adv. Space Res. 2, 241

\bibitem{} Unger S.W., Pedlar A., Hummel E., 1989, A\&A 208, 14 

\bibitem{} van Driel W., van Woerden H., 1994, A\&A 286, 395

\bibitem{} Weaver K.A., Wilson A.S., Henkel C., Braatz J.A., 1999, to appear in 
           ApJ; astro-ph/9902269 

\bibitem{} Wiklind T., Henkel C., 1990, A\&A 227, 394 

\bibitem{} Wiklind T., Combes F., Henkel C., 1995, A\&A 297, 643 

\bibitem{} Wills D., 1967, ApJ 148, L57

\bibitem{} Wills D., Wills B.J., 1976, ApJS 31, 143

\bibitem{} Wisotzki L., Bade N., 1997, A\&A 320, 395

\bibitem{} Zasov A.V., Sil'chenko O.K., 1987, Pis'ma Astron. Zh. 13, 455


\bibitem{} Zimmermann H.U., Becker W., Belloni T., et al., 1994,
               MPE Report 257

\end{thebibliography}
\end{document}